\documentclass[3p,times]{elsarticle}

\usepackage{ecrc}
\volume{00}
\firstpage{1}
\journalname{Nuclear Physics A}
\runauth{A. Buzzatti and M. Gyulassy}
\jid{nupha}

\usepackage{amssymb}
\usepackage[figuresright]{rotating}
\usepackage{wrapfig}

\newcommand{\be}{\begin{equation}}
\newcommand{\ee}{\end{equation}}
\newcommand{\bea}{\begin{eqnarray}}
\newcommand{\eea}{\end{eqnarray}}
\newcommand{\lp}{\left(}
\newcommand{\rp}{\right)}
\newcommand{\bk}{\mathbf{k}}
\newcommand{\bq}{\mathbf{q}}
\newcommand{\bx}{\mathbf{x}}

\begin{document}

\begin{frontmatter}
\title{Dynamical magnetic enhancement of light and heavy quark \\jet quenching at RHIC}
\author{Alessandro Buzzatti and Miklos Gyulassy}
\address{Department of Physics, Columbia University, 538 West 120th Street, New York, NY 10027, USA}
\begin{abstract}
A new Monte Carlo implementation of Djordjevic's dynamical scattering generalization of the DGLV radiative energy loss opacity series is used with a hybrid interpolation scheme to compute both light and heavy quark jet quenching up to third order in opacity. The enhancement of the ratio of bottom to charm quark energy loss due to perturbative long range color magnetic effects in nonuniform Bjorken expanding geometries is 
found to reduce the significance of  the heavy quark jet puzzle posed by the observed near equality (within sizeable errors) of pion and  nonphotonic electron nuclear modification at RHIC. Jet Flavor Spectroscopy discussed below will be a powerful tool to differentiate competing dynamical models of the QGP produced in ultra-relativistic nuclear collisions.
\end{abstract}
\end{frontmatter}

\section{Introduction}

We report results of a recent detailed Monte Carlo study\cite{us} of dynamical magnetic scattering effects on the radiative energy loss of light and heavy quark jets in Au+Au collisions at $\sqrt{s}=200$ AGeV. The aim of this work is to advance the theory of jet tomography for quantitative applications to heavy ion collisions in accord with the objectives of the DOE topical JET collaboration project\cite{JETcollab0}.

Since 2005 the heavy quark jet puzzle\cite{RAAe05} has remained\cite{MG09} one of the outstanding unsolved problems within the perturbative QCD multiple collision theory framework.  Predictions of heavy quark quenching based on the DGLV opacity series model\cite{ref4} and other pQCD models of radiative energy loss were found to significantly over-predict the midrapidity nonphotonic electron yield in the $p_T\sim 5$ GeV range due to the  relatively small energy loss of heavy bottom quark jets. Even after inclusion in WHDG\cite{ref5} of additional elastic energy loss and fluctuations of path lengths effects (as required to maintain consistency of the theory with the  observed light parton/pion nuclear modification factor $R_{AA}^\pi\sim 0.2$),  the bottom quark contribution to the nonphotonic electron spectrum was found to be too weakly quenched to explain the PHENIX and STAR data\cite{RAAe05}. This discrepancy has led to novel gravity dual holographic models of jet energy loss completely outside the pQCD framework (see \cite{MG09, Horowitz:2007su, NGT10} and refs. therein).

In this work we focus only on recent refinements of perturbative Hard Thermal Loop (HTL) QCD dynamical theory that have not been quantitatively studied up to now. In particular, we follow the proposal of (MD) M.Djordjevic \cite{ref6} that long range dynamical color magnetic scattering in the HTL framework may significantly enhance the bottom quark radiative energy loss and thereby possibly help solve the heavy quark puzzle without having to invoke new non-perturbative dynamical assumptions. We have therefore developed a Monte Carlo hybrid interpolation version of the DGLV-BFW-MC1.0 code\cite{BFW1}  to evaluate numerically the sensitivity of the ratio of charm and  bottom quark jet quenching to perturbative color magnetic effects up to moderate order in opacity.

The hybrid DGLV-BFW-MC approach, used here, involves replacing the GW\cite{Gyulassy:1993hr} ({\em static} Debye screened) momentum distributions in the GLV(eq.113)/DGLV(eq.17) multiple collision kernel 
$$
\Pi_{i=1}^N dz_id^2{\bf q_i} \sigma(z_i) \rho(z_i) 
[\bar{v}^2(z_i,{\bf q}_i) -\delta^2({\bf q}_i))]$$
with a path dependent  effective magnetic enhanced transverse distributions
\be
	 \bar{v}^2(z,{\bf q}; r_m)= 
\frac{\mu_e(z)^2 {\cal N}(r_m)}{\pi}
\frac{1}{(\lp \bq^2 + \mu_e^2(z)
             \rp)(\lp \bq^2 + r_m^2 {\mu}_e^2(z) \rp)}
\;\;\;\;,\ee
where $0\le r_m\equiv \mu_m/\mu_e \le 1$ is the ratio of the color electric Debye and the assumed longer color magnetic screening lengths. The normalization factor ${\cal N}(r)=(1-r^2)/Log[1/r^2]$ reduces to unity when $r=1$ but has a weak logarithmic zero for $r\rightarrow 0$. This zero is however cancelled  by the weak divergence of the effective elastic cross section $\sigma(z,r_m)=9\pi\alpha^2/(2 \mu_e^2(z){\cal N}(r_m))$ as shown in detail by Djordjevic\cite{ref6} for $N=1$. However, in higher opacity orders the unitarity corrections diverge in the (unphysical) $r_m=0$ limit and require more care. Fortunately, in the finite size quark gluon plasma produced in A+A, $r_m$ is bounded from below by $1/(\mu_e R_A)$ due to confinement of color outside the finite size plasma. We consider here finite $r_m \ge 1/3$ to explore the sensitivity of the bottom to charm jet ratio to enhanced soft momentum transfer due to perturbative magnetic field fluctuations of range up to $3/gT$.

\section{Static Brick Problem}

At first order in opacity, the MD ($r_m=0$) model\cite{ref6} of the induced gluon number radiated per light cone momentum fraction from a massive quark jet produced at position ${\bf x}$ with energy $E$ in azimuthal direction $\phi$ is given by
\be
	\frac{dN_g}{dx_+}(\bx,\phi) = \frac{C_R \alpha_s}{\pi}  
	\int d\tau \frac{d^2k}{\pi} \frac{d^2q}{\pi}
	\frac{1}{x_+} \frac{\frac{9}{2}\pi\alpha^2}{\bq^2 (\bq^2{+}\mu^2)}
  \frac{2(\bk{+}\bq)}{(\bk{+}\bq)^2{+}\chi}
	\lp \frac{(\bk{+}\bq)}{(\bk{+}\bq)^2{+}\chi} - \frac{\bk}{\bk^2{+}\chi} \rp
  \lp 1- \cos\lp\frac{(\bk{+}\bq)^2+\chi}{2x_+E} \tau\rp \rp \rho_{QGP}(\bx,\tau)
   \;\;,
\ee
where $x_+$ is the fraction of plus momentum carried away by the radiated gluon, $\mu=gT({\bf x},\phi,\tau)$ is the local path dependent Debye screening mass and $\lambda^{-1}_{dyn}=3\alpha_s T({\bf x},\phi,\tau)$ is the inverse local dynamical mean free path. Here $\chi=M^2 x_+^2 + m_g (1-x_+)$ controls the ``dead cone'' effect due to the finite jet quark  mass $M$ as well as the local effective gluon thermal mass $m_g=\frac{\mu({\bf x},\phi,\tau)}{\sqrt{2}}$. For the idealized  static ``brick'' problem $\rho_{QGP}=\theta(L-\tau)$. For $r_m>0$ we use our interpolation distribution eq.(1). We include effects of Poisson fluctuations of the radiative energy loss due to gluon number fluctuations to compute $P(\epsilon)$, the probability distribution of radiating a fraction of energy $\epsilon$.

\begin{figure*}[!h]
\label{dEL}
	\includegraphics[height=0.35\textwidth]{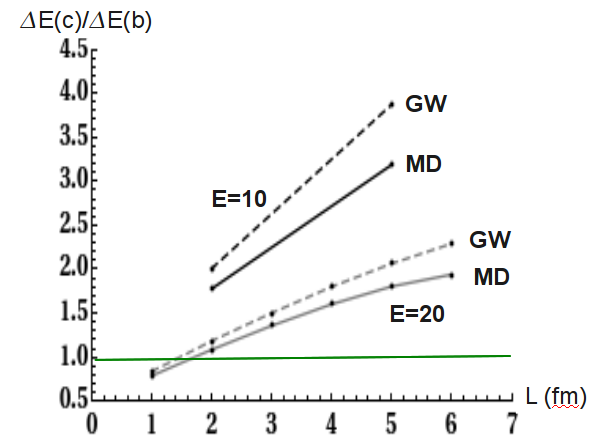}
	\includegraphics[height=0.35\textwidth,width=0.45\textwidth]{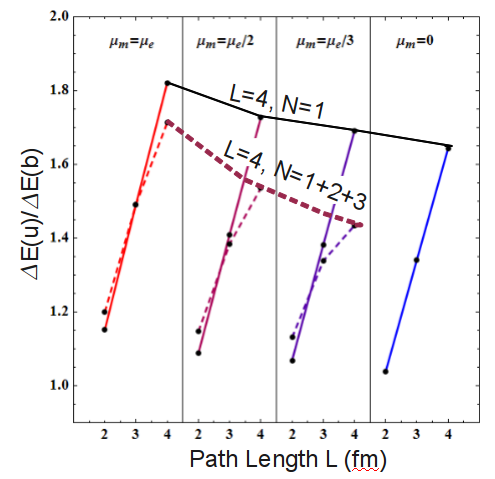}
	\caption{\small{\textit{Left: The ratio of charm to bottom jet radiative energy loss versus path length in a uniform static QGP at $T=250MeV$ is shown for $E=10$ and $20$ GeV for the static GW\protect{\cite{Gyulassy:1993hr}} distribution ($r_m=1$ in eq.1) compared to the dynamic MD\protect{\cite{ref6}} distribution ($r_m=0$). Right:  The ratio of light up quark (thermal mass) to heavy bottom quark energy loss vs L for $E=20$ GeV jets and different $r_m=1,1/2,1/3,0$. Solid lines correspond to $N=1$ while dashed lines correspond to hybrid MC results with $N=1+2+3$. The ideal $r_m=0$ case is the MD model\protect{\cite{ref6}}. }}}
\end{figure*}

The left panel of Fig.1 shows that the ratio of charm to bottom energy loss at first order in opacity is reduced by approximately 20\% in the dynamical MD\cite{ref6} compared to the static GW\cite{Gyulassy:1993hr} case, but the ratio for $L>2$ fm remains well above unity in both limits. In the right panel the ratio of light (up) quark energy loss to bottom quark energy is shown vs path length $2\le L\le 4$ fm for different $r_m$ and different orders in opacity. It is found that charm and up quarks have essentially identical energy loss. What is clear is that neither long range magnetic scattering nor higher orders are sufficient to bring the ratio near unity for $L>2$ fm.
 
Despite the fact that the absolute value of $\Delta E(Q)$ is almost doubled by switching from the pure static $r_m=1$ to the dynamical $r_m=0$ model in the uniform brick geometry, as seen in Fig.2, the c/b and u/b ratios in Fig.1 are surprising insensitive to the magnetic screening length.

\section{Bjorken expansion}

To test the effect of magnetic enhanced scattering rates in a more realistic diffuse transverse geometry with longitudinal Bjorken expansion cooling we assumed Wood-Saxon participant density profile and $\tau_0/\tau$ proper time dependence for $\rho_{QGP}({\bf x},\tau)$ in eq.2. We terminate the time integral when the Debye screening mass drops below $\Lambda_{QCD}$. Expansion reduces the effective path length and this helps to reduce the ratio of charm to bottom energy loss as seen from Fig.1. We average initial jet production points according to the Glauber $T_{AA}({\bf x},b)$ profile and average over the jet $\phi$ angles.

Analogously to the brick setup, for each of the jets we compute the probability distribution $P(\bx,\phi;\epsilon)$ and then average the results over the initial jet production profile in order to obtain the radiative energy loss $\Delta E /E$ and the partonic $R_{AA}$:
\be
\label{Raa}
	\frac{\Delta E}{E} = \int \frac{d\phi}{2\pi} d\bx \rho_{Jet}(\bx) \int d\epsilon \epsilon P(\bx,\phi;\epsilon) \ \ \ \ \ \ \ \ \
	R_{AA} \approx \int \frac{d\phi}{2\pi} d\bx \rho_{Jet}(\bx) \int d\epsilon (1-\epsilon)^n P(\bx,\phi;\epsilon) \;\;.
\ee

The approximation in (\ref{Raa}) is strictly valid only if the initial Au+Au spectrum is equal to the p+p spectrum and proportional to a constant index power law $n$, here set to $4$. However, it provides a simple estimate that allows us to compare the MD and DGLV models in a static uniform cylindric geometry or a Bjorken expanding one. The results are shown in Fig 2, where we plot ${\Delta}E(M)/E$ and $R_{AA}(M)$ as a function of the mass of the quark for $E=20 GeV$.
\begin{figure*}[!h]
	\label{Bj}
\includegraphics[height=0.4\textwidth]{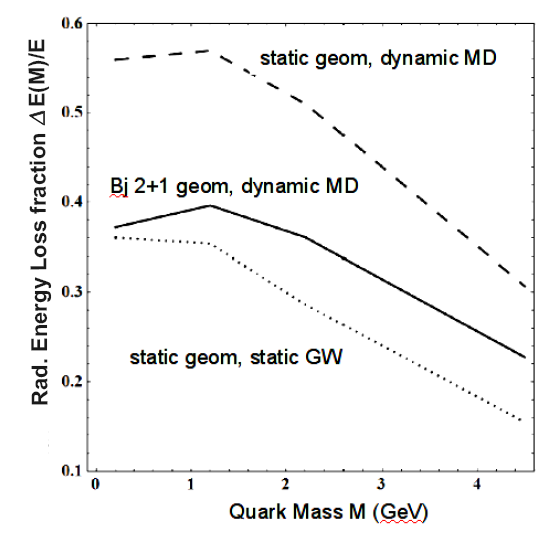}
\includegraphics[height=0.4\textwidth]{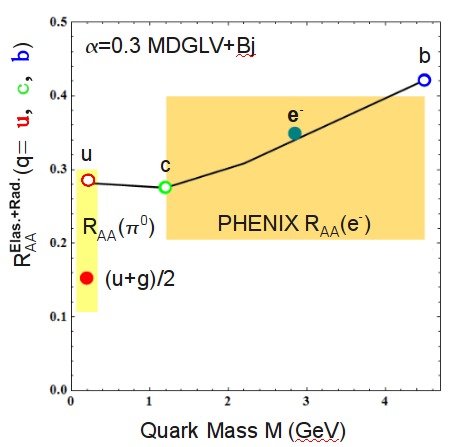}
	\caption{\small{\textit{Illustration of Jet Flavor Spectroscopy. Left: Comparison between GW\protect{\cite{Gyulassy:1993hr}} and MD\protect{\cite{ref6}} average radiative energy loss ${\Delta}E(M)/E$ as a function of $M$, for uniform cylindric and Bjorken expanding geometries. Right: radiative plus elastic plus gluon fluctuation $R_{AA}(M)$ vs parton flavor mass is shown for Bjorken expanding diffuse Glauber geometry including MD magnetic enhancement and $\alpha_s=0.3$. The empty circles represent the partonic contribution of the up,charm,bottom quarks. The filled circles are estimates of the pion and electron $R_{AA}$ assuming equal contribution of light quarks/gluons and charm/bottom jets respectively. The broad rectangular shaded box represents the current PHENIX and STAR data. Elimination of the wide horizontal uncertainty bar in $R_{AA}(e^-)$ requires future c and b jet flavor tagged measurements.}}}
\end{figure*}

The left panel of Fig. 2 shows  that the dynamical MD scattering increases the energy loss fraction in a static plasma by about a factor of two. However, after including the dilution effects due to the expanding geometry, ${\Delta}E(M)/E$ decreases back toward the uniform static geometry, static GW value. Also quite remarkably, there is an approximate quark flavor independence from $0< M < 2$ GeV. Thus, identified charm quark jets can serve as calibrated light quark jet probes and hence can serve to uncover the gluon jet contribution to the pion $R_{AA}$. Regarding the bottom quark, we find a modest increase of the energy loss with respect to GW and static geometries, but not sufficient to enhance $\Delta E(b)/\Delta E(c)$ beyond about 0.5.

On the right panel, $R_{AA}(Q)$ is shown after adding the mean elastic energy loss contribution from \cite{ref5} for $\alpha=0.3$. The figure provides a convenient and intuitive ``Jet Flavor Spectroscopy'' representation of theory as well as experiment. It (1) emphasizes that pQCD tomography with light (u,d,s) and charm quark jets is essentially equivalent, and thus tagged charm tomography fixes the light quark component of the high pT pion quenching $R_{AA}(\pi)\approx f_g R_{AA}(g)+ f_{uds} R_{AA}(c)$ and (2) shows the mass splitting of $R_{AA}$ between b and c quark jets which depends on both the detailed pQCD dynamics as well the diffuse expanding QGP geometry.

The shadowed regions represent the current rather large error bars in the pion and nonphotonic electron data. Once we take into account such large error bands, the inclusion of color magnetic effects in diffuse expanding geometries does help to reduce the significance of the heavy quark puzzle as posed in WHDG\cite{ref5}. In this representation, the jet flavor spectroscopy hierarchy predicted in the pQCD framework including dynamical scattering as well as dynamical geometry $$R_{AA}(g) < R_{AA}(\pi \sim (g+u)/2) < R_{AA}(c) <R_{AA}(e^- \sim (c+b)/2)) <R_{AA}(b)<R_{AA}(\gamma)$$ may actually not be incompatible with present light and heavy jet tomographic data.

\section{Conclusions}
Our main conclusion is that flavor tagged charm and bottom quark jet data is essential to confirm  or reject the predicted pQCD nuclear modification flavor hierarchy that can be conveniently revealed using the ``Jet Flavor Spectroscopy'' diagram Fig.2b. The mass splitting between c and b jets is a particularly robust observable to differentiate pQCD and holographic gravity dual modelling of jet-plasma interactions, as emphasized in \cite{Horowitz:2007su}. Our preliminary results with hybrid Monte Carlo and $1/3< r_m=\mu_m/\mu_e <1$ up to third order in opacity indicate that magnetic effects help drive $R_{AA}$ down toward the present RHIC data and also reduce the b/c mass splitting. However, this hybrid pQCD approach still predicts that $R_{AA}(e^-)/R_{AA}(\pi)\approx 2$ due to the highly quenched gluon jet component of the $\pi$ yield \cite{ref5}. Once $R_{AA}(c)$ is known, the gluon $R_{AA}(q)$ also becomes experimentally accessible since $c\sim u$ in this framework. As a final point, we assumed no initial state saturation or shadowing effects. Those have to be determined from $p+A$ or direct photon $R_{AA}(\gamma)$ to allow deconvolution of  initial from final state quenching effects.    

{Acknowledgements}: We thank Andrej Ficnar, Simon Wicks, and Magdalena Djordjevic for helpful discussions in the course of this work. We acknowledge support by US-DOE Nuclear Science Grant No. DE-FG02-93ER40764.

\bibliographystyle{elsarticle-num}

\end{document}